\documentclass[twocolumn,preprintnumbers,amsmath,amssymb,superscriptaddress]{revtex4}

\usepackage{graphicx}
\usepackage{dcolumn}
\usepackage{braket}
\usepackage{color}

\begin{document}

\title{Predissociation dynamics of lithium iodide}

\author{H. Schmidt}
\affiliation{Physikalisches Institut, Universit{\"a}t Freiburg, 
  79104 Freiburg, Germany}

\author{J. von Vangerow}
\affiliation{Physikalisches Institut, Universit{\"a}t Freiburg, 
  79104 Freiburg, Germany}

\author{F. Stienkemeier}
\affiliation{Physikalisches Institut, Universit{\"a}t Freiburg, 
  79104 Freiburg, Germany}

\author{A. S. Bogomolov}
\affiliation{Institute of Chemical Kinetics and Combustion, 
  Novosibirsk 630090, Russia}

\author{A. V. Baklanov}
\affiliation{Institute of Chemical Kinetics and Combustion, 
  Novosibirsk 630090, Russia}
\affiliation{Novosibirsk State University, Novosibirsk 630090,
  Russia}

\author{D. M. Reich}
\affiliation{Theoretische Physik, Universit{\"a}t Kassel, 
  Heinrich-Plett-Str. 40, 34132 Kassel, Germany}

\author{W. Skomorowski}
\affiliation{Theoretische Physik, Universit{\"a}t Kassel, 
  Heinrich-Plett-Str. 40, 34132 Kassel, Germany}

\author{C. P. Koch}
\affiliation{Theoretische Physik, Universit{\"a}t Kassel, 
  Heinrich-Plett-Str. 40, 34132 Kassel, Germany}

\author{M. Mudrich}
\email{mudrich@physik.uni-freiburg.de}
\affiliation{Physikalisches Institut, Universit{\"a}t Freiburg, 
  79104 Freiburg, Germany}

\date{\today}


\begin{abstract}
  The predissociation dynamics of lithium iodide (LiI)
  in the first excited $A$-state is investigated for molecules in the
  gas   phase and embedded in helium nanodroplets, using
  femtosecond pump-probe photoionization spectroscopy. In the gas
  phase, the transient Li$^+$ and LiI$^+$ ion signals feature damped
  oscillations due to the excitation and decay of a vibrational wave
  packet. Based on high-level {\it ab initio} calculations of the
  electronic structure of LiI and simulations
  of the wave packet dynamics, the exponential signal decay is found
  to result from predissociation predominantly at the lowest avoided
  $X$-$A$ potential curve crossing, for which we infer a coupling constant
  $V_{XA}=650(20)$ cm$^{-1}$. The lack of a pump-probe
  delay dependence for the case of LiI embedded in helium nanodroplets
  indicates fast droplet-induced relaxation of the vibrational
  excitation.  
\end{abstract}

\maketitle

\section{Introduction}\label{sec:Intro}

Three decades ago, the dream of controlling chemical reactions with
laser pulses has initiated the field of coherent
control~\cite{TannorJCP85,TannorJCP86,BrumerShapiroCPL86}. The idea is
to selectively control the outcome of photo-induced reactions -- formation of a
chemical bond or creation of photofragments -- by changing the
parameters of the laser pulses.
Femtosecond pump-probe spectroscopy has been instrumental in bringing
this dream closer to reality, terming the notion 
``femtochemistry''~\cite{Zewail:1988,Zewail:1994,Manz:1995,Chergui:1995}. 

Alkali halide diatomics were among the first molecules to be studied by
femtosecond spectroscopy in the gas phase. Their lowest electronic
excitations mostly lie in the visible or ultraviolet spectral range
and are therefore well accessible by femtosecond lasers.  
Alkali halides are special in that their chemical bond has ionic
character even in the electronic groundstate. This character results from the
extremely differing electron affinities of the constituent atoms. As a
consequence, even the lowest excited states are subject to curve
crossings with the ion-pair state potential curve at intermediate
distances between the nuclei. These curve crossings give rise to
non-adiabatic couplings and to predissociation of the photo-excited
molecules.  

In pioneering time-resolved experiments, Zewail and coworkers have
visualized the predissociation dynamics of sodium iodide (NaI) and
other alkali halides as the repeated passage of a coherent vibrational
wave packet across the avoided curve crossing of the groundstate and
the first excited state~\cite{Rose:1988,Rose:1989,Mokhtari:1990}. At
every passage the wave packet splits up into one fraction which
remains in a bound state and continues to vibrate, and another fraction
which follows the dissociative branch of the potential, 
producing unbound Na and I
fragments~\cite{Su:1979,Rose:1988,Rose:1989,Moller:2000}. 

Such textbook quantum molecular dynamics can typically only be
observed in gas phase experiments where the molecules are
isolated from their environment. In the condensed phase,
the coherent motion induced by a femtosecond laser pulse is most often
subject to very fast decay due to relaxation and dephasing processes. 
A noticable exception present diatomic alkali metal molecules attached
to helium
nanodroplets~\cite{Claas:2006,Claas:2007,Schlesinger:2010,Mudrich:2009,Gruner:2011}. 
Helium (He) droplets are quantum fluid clusters which feature
extraordinary properties such as microscopic superfluidity~\cite{Grebenev:1998}, a critical velocity 
for the undamped motion of embedded molecules~\cite{Brauer:2013}, and quantized 
states of angular momentum (vortices)~\cite{Gomez:2014}.
The dynamical response of He droplets can be probed by the wave packet motion of impurities such as alkali dimers. 
The vibrational dynamics of these molecules was found to be only weakly
perturbed by the droplets. This can be explained by a weak coupling
of alkali metal atoms and small molecules to helium droplets due to
their position at the droplet surface. In contrast, recent studies of
the impulsive alignment of molecules embedded in the droplet interior
have evidenced that the rotational wave packet dynamics is
significantly slowed down and rotational recurrences are completely suppressed~\cite{PentlehnerPRL:2013}.

Here, we present a femtosecond pump-probe
photoionization study of free LiI and of LiI embedded in helium nanodroplets. Lithium iodide, as all other
molecules which do not purely consist of alkali or alkaline earth
metals, are immersed in the interior of He droplets. Therefore
strong perturbations of vibrational dynamics by the He environment may
be expected. We start by investigating 
the photodynamics of LiI in the gas phase which 
is much less well characterized than that of
NaI. The latter has been studied in detail, both experimentally using
femtosecond spectroscopy, and theoretically using semiclassical and quantum
dynamical model calculations~\cite{Rose:1988,Mokhtari:1990,Cong:1996,Jouvet:1997,Engel:1989,Braun:1996,Chapman:1991,Arasaki:2003}. 
Preliminary results on LiI~\cite{Rose:1989} suggest a similar
predissociation dynamics as for NaI. To clarify whether this
expectation is indeed correct, we perform
one-color femtosecond pump-probe
measurements in the UV spectral range. We corroborate the
experimental results with high-level {\it ab initio}
calculations to determine the potential energy curves and
non-adiabatic couplings from first principles. Based on this data, we 
model the photoionization dynamics of LiI, using both semiclassical
arguments and full quantum dynamical simulations. 

Our paper is organized as follows. Section~\ref{sec:pot} introduces
the electronic structure of LiI, outlining first the excitation scheme
and discussing then the details of the electronic structure
calculations. The experimental setup and results are presented in
Sec.~\ref{sec:exp} for gas phase measurements
(Sec.~\ref{subsec:GasResults}) and droplet 
experiments (Sec.~\ref{subsec:droplets}). We compare the experimental
data to simulation results in Sec.~\ref{sec:Simulation}. Our
findings are summarized in Sec.~\ref{sec:summ}.



\section{Electronic structure of L\MakeLowercase{i}I}
\label{sec:pot}

\subsection{Excitation scheme}
\label{subsec:scheme}

\begin{figure}
\centering
\includegraphics[width=0.4\textwidth]{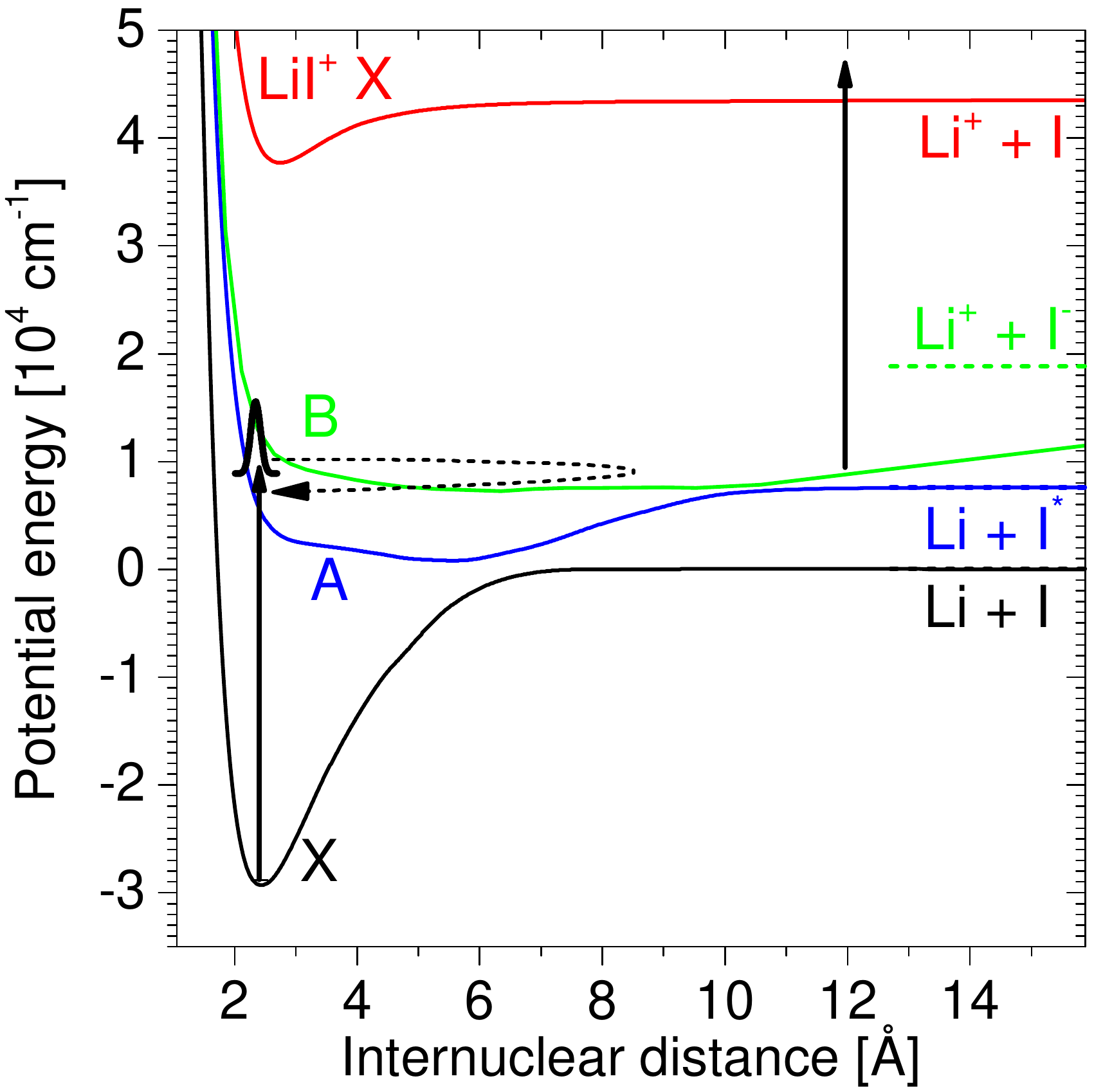}
\caption{Pump-probe scheme employed in the present experiments. The
  pump pulse excites a vibrational wave packet out of the
  groundstate $X$ into the first excited state $A$. The probe
  pulse ionizes LiI by exciting into the groundstate of LiI$^+$ or
  into the Li$^+$ + I dissociative continuum
  near the outer turning point of the $A$-state.} 
\label{fig:scheme}
\end{figure}
The excitation and ionization scheme of LiI is depicted in
Fig.~\ref{fig:scheme} using the adiabatic potential energy curves
obtained by \textit{ab initio} calculations as described below. 
Analogously to NaI and other alkali halides, the structure of
adiabatic potential energy curves 
is determined by the low-lying Li$^+$-I$^-$ ion-pair state which
crosses the covalent states at intermediate internuclear distances.
This causes the groundstate $X$ to have a deep potential well
($D_0=28839\,$cm$^{-1}$, $R_e=2.39\,$\AA~\cite{Su:1979}) and the first
excited state $A$ to be predissociative. The crossing of the $A$-state
potential with the ionic potential at
$R_{x,A}=6.13\,$\AA~\cite{Schaefer:1986} is avoided due to
non-adiabatic as well 
as spin-orbit couplings with a coupling energy $V_{XA}$ ranging
between 686$\,$cm$^{-1}$ and 766$\,$cm$^{-1}$ according to the
literature~\cite{Delvigne:1973,Grice:1974,Schaefer:1986}. A
second curve crossing exists due to the intersection of the
$B$-state potential with the ionic Li$^+$-I$^-$ potential around
$R_{x,B}= 10\,$\AA~where $V_{AB}\approx 5$ cm$^{-1}$ according to our
calculations. To the best of our knowledge, no literature data are
available on the $B$-state. 

A vibrational wave packet, excited in the $A$-state by a first
ultrashort laser pulse (left vertical arrow in Fig.~\ref{fig:scheme}), 
oscillates back and forth from the inner to the outer classical
turning points. Due to the avoided $X$-$A$ curve crossing, part of the
wave packet escapes outwards along the dissociative diabatic potential
to form free Li + I fragments each time the wave packet comes near the
crossing point. Since the ionization energy of atomic Li ($5.39\,$eV)
exceeds the photon energy ($\sim 4.6\,$eV), only the bound excited
molecule LiI$^*$ can be ionized by one-photon absorption from a second
probe pulse (right vertical arrow). The fraction of
the wave packet which follows the adiabatic potential curve
converging toward the ionic curve at large $R$ can be ionized by
photodetachment of the I$^-$ component, leaving behind
Li$^+$+I. Ionization is enhanced at the outer turning point
of the wave packet oscillation around $R_i=12\,$\AA~due to
accumulation of wave packet amplitude at $R_i$, which corresponds to
a maximum classical dwell time. 
At short laser wavelengths
$\lambda\lesssim 270\,$nm the $B$-state is also partly populated. This 
leads to fast dissociation along the diabatic $B$-state potential
curve.

\subsection{\textit{Ab initio} calculations}
\label{subsec:abinitio}

The electronic structure of the LiI molecule, briefly introduced in
the previous section, is determined in state of the art  \textit{ab
  initio} calculations to allow for a quantitative comparison
between theory and experiment. To ease calculations, the LiI molecule is
treated as a system with effectively 10 electrons. This is achieved by 
accounting for the core electrons of the iodide atom by a
pseudopotential. To account for relativistic effects, 
the ECP46MDF pseudopotential from the Stuttgart
library~\cite{Stoll:2002} is employed for the core electrons, leaving
7 valence electrons  ($5s^25p^5$) in I which are explicitly treated in
the calculations. For the I valence
electrons we adopt the standard pseudopotential basis set [$6s6p/4s4p$]
extended by [$13d3f2g/5d3f2g$] functions taken from the
quadrupole-quality basis set suggested in
Ref.~\cite{Peterson:2006}. For the Li atom the aug-cc-pVQZ basis
set is employed~\cite{Dunning:1989}.  
The {\it ab initio} calculations of the LiI  potentials described in
the following are performed using the MOLPRO package~\cite{Molpro:2010}. 

\begin{figure}
\centering
\hspace*{-4em}
\includegraphics[width=0.45\textwidth]{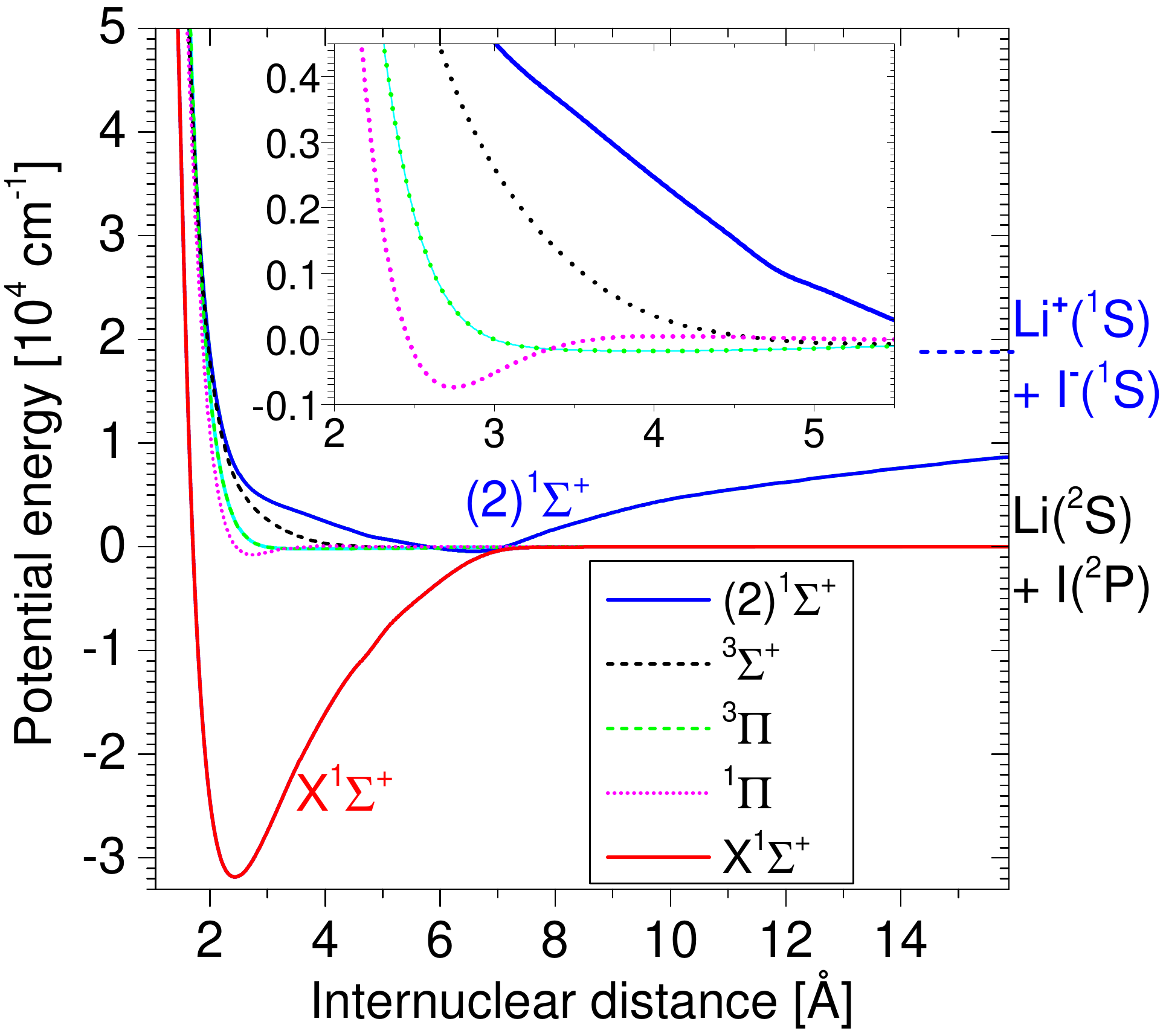}
\caption{Calculated potential energy curves for the low-lying
  non-relativistic ($S-\Lambda$) states of the LiI
  molecule. The inset shows a close-up view of the
    potentials at short distance.} 
\label{fig:Non_rel_curves}
\end{figure}
In the first step we calculate the lowest nonrelativistic ($S-\Lambda$) states
$X^1\Sigma^+$, $(2)^1\Sigma^+$, $^1\Pi$, $^3\Sigma^+$, and $^3\Pi$,
see Fig.~\ref{fig:Non_rel_curves}. 
These states asymptotically correlate with the groundstate atoms,
Li($^2S$)+I($^2P$), and with the ionic asymptote, Li$^+$($^1S$)+I$^-$($^1S)$.
The nonrelativistic singlet (low-spin) states, $X^1\Sigma^+$,
$(2)^1\Sigma^+$, and $^1\Pi$, have been calculated using the
internally contracted multireference configuration interaction method
with  single and double excitations (MRCI)~\cite{Werner:1988} with
additional Davidson correction (MRCI+Q)~\cite{Langhoff:1974}. The MRCI
calculations are preceded by state-averaged complete-active-space
calculations (CASSCF)~\cite{Werner:1985} to generate optimized
orbitals and reference state energies for the MRCI runs. The
multireference method is essential here since we describe low-spin
electronic states which asymptotically correlate with two open-shell
atoms, $^2S$ + $^2P$. The Davidson correction is highly important for
the final results since it largely compensates for the lack of
size-extensivity of the MRCI method.

For the triplet states of LiI, $^3\Sigma^+$ and $^3\Pi$, and for the
lowest states of the LiI$^+$ ion, $^2\Sigma$ and  $^2\Pi$, we could
employ the open-shell spin-restricted coupled cluster method with
single, double and non-iterative triple excitations,
RCCSD(T)~\cite{Knowles:1993}, which is a fully size-consistent
method. The single reference coupled cluster method can be used here
since these are the lowest high-spin states in a given symmetry. 
The long-range tail of the high-spin states, $^3\Sigma^+$ and $^3\Pi$,
can easily be obtained from the supermolecular RCCSD(T)
calculations. However, the MRCI+Q results would not yield reliable
interaction energies at long range and we need an alternative
approach for low-spin states. For the ionic $(2)^1\Sigma^+$ state, the
long-range part of the potential ($R > 13\,$\AA ) has  been assumed to
be pure  $-1/R$ Coulomb interaction. For the low-spin singlet
states,  $X^1\Sigma^+$ and $^1\Pi$, we have assumed the long-range
interaction energy to be equal to the long-range part of $^3\Sigma^+$
and $^3\Pi$ states, respectively. This is a well justified
approximation, since we know that van der Waals coefficients $C_{n}$
($n\ge6$) depend on the asymptotic atomic states but not on the
resultant molecular spin; they thus are exactly the same for the low
and high-spin states dissociating into the same atoms. 
The spin-orbit couplings between the nonrelativistic ($S-\Lambda$)
states are calculated using the MRCI method with the spin-orbit
Hamiltonian $\hat{H}_{{\rm SO}}$ in the Breit-Pauli approximation. The 
nonadiabatic radial coupling between the $X^1\Sigma$ and the
$(2)^1\Sigma$ states is also calculated by employing the MRCI
method. 
 
Finally we diagonalize the matrix $\hat{H}_{{\rm nrel}} +
\hat{H}_{{\rm SO}}$, where the elements of the diagonal $\hat{H}_{{\rm
    nrel}}$ matrix come from the MRCI+Q (low-spin states) or
RCCSD(T)(high-spin states) calculations. After including
spin-orbit couplings, the electronic states do not have any more a 
defined spin $S$ or projection of the orbital angular momentum
$\Lambda$ on the molecular axis. The only conserved quantum number is
now the projection of the total electronic angular momentum (spin +
orbital) on the molecular axis, denoted by $\Omega$.  
The $\hat{H}_{{\rm nrel}} + \hat{H}_{{\rm SO}}$  matrix is
block-diagonal, and each block can be labeled by a  different quantum
number $\Omega$. From the set of states we consider, $X^1\Sigma^+$,
$(2)^1\Sigma^+$, $^1\Pi$, $^3\Sigma^+$, and $^3\Pi$, we obtain four
blocks with $\Omega = 0^+, 0^-, 1$ and 2. Diagonalization of each of
these blocks yields the adiabatic states with given $\Omega$. For our
purposes, the most interesting is the block with symmetry $0^+$, which
includes the ground adiabatic state $X0^+$. Our model yields three
$0^+$ states: $X0^+$ , $A0^+$, and $B0^+$, which asymptotically
correlate to Li($^2S$)+I($^2P_{3/2}$),  Li($^2S$)+I($^2P_{1/2}$) and
Li$^+$($^1S$)+I$^-$($^1S$), respectively.

The two $^1\Sigma^+$ states (or $X0^+$ and $A0^+$ after including the
SO-coupling), strongly interact with each other near the
avoided-crossing. To account for that, we calculate the nonadiabatic
radial coupling between the two states, $\tau(R)$, as a function of
internuclear distance, $R$, 
\begin{equation*}
  \tau(R)=  \langle X^1\Sigma^+ |\frac{d}{d R}| (2)^1\Sigma^+ \rangle.
\end{equation*}
The coupling $\tau(R)$ is then used to calculate the mixing angle
$\gamma(R)$, 
\begin{equation*}
  \gamma(R)= \int_{R}^{\infty} \tau(s)  \rm{d} s.
\end{equation*}
In the next step, the mixing angle $\gamma(R)$ is employed to make a 
transformation from the adiabatic to the diabatic representation, 
\begin{eqnarray*}
  V_{1}^{\rm d}(R) &=&V_{2}^{{\rm ad}}(R)\sin^2\gamma(R)
  + V_{1}^{{\rm ad}}(R)\cos^2\gamma(R),\\
  V_{2}^{\rm d}(R) &=&V_{1}^{{\rm ad}}(R)\sin^2\gamma(R)
  + V_{2}^{{\rm ad}}(R)\cos^2\gamma(R),\\
 V_{12}^{\rm d}(R) &=&(V_{2}^{{\rm ad}}(R)
 -V_{1}^{{\rm ad}}(R))\sin\gamma(R) \cos\gamma(R),
  \label{adiab}
\end{eqnarray*}
where $V_{1}^{{\rm ad}}(R)$ and $V_2^{{\rm ad}}(R)$ are the adiabatic
$X^1\Sigma^+$ and $(2)^1\Sigma^+$ potentials, respectively,
$V_{1}^{\rm d}(R)$ and  $V_{2}^{\rm d}(R)$ are the diabatic states,
and $V_{12}^{\rm d}(R)$  is the diabatic coupling potential. This
diabatization procedure enables to eliminate the  derivative with
respect to $R$, $\tau(R) \;\cdot\; d/d R$, from the equations
which describe the nuclear motion~\cite{Baer:2002}. In the dynamical
calculations this derivative  is  replaced  by the coupling potential
$V_{12}^{\rm d}(R)$. It is highly advantageous  since $\tau(R)$ is a
strongly varying function of $R$ (approaching the Dirac delta form
near the avoided-crossing), while $V_{12}^{\rm d}(R)$ is a smooth
function of $R$.
In Hund's case $(a)$ representation, the Hamiltonian governing the
vibrational dynamics becomes
\begin{equation}
  \hat{H} = 
  \begin{pmatrix} 
    \hat{T}+V_{1}^{\rm d}(R) & V_{12}^{\rm d}(R) & 
    \langle V_{1}^{\rm d}|\hat{H}_{{\rm SO}}| ^3\Pi\rangle \\ 
    V_{12}^{\rm d}(R) & \hat{T}+V_{2}^{\rm d}(R) & 
    \langle V_{2}^{\rm d}|\hat{H}_{{\rm SO}}|^3\Pi\rangle \\
    \langle V_{1}^{\rm d}|\hat{H}_{{\rm SO}}|^3\Pi\rangle & 
    \langle V_{2}^{\rm d}|\hat{H}_{{\rm SO}}|^3\Pi\rangle& 
    \hat{T}+\frac{1}{3}\Delta+V^{^3\Pi}(R) 
  \end{pmatrix}
  \label{hampot}
\end{equation}
where $\hat{T}$ is the kinetic operator and $\Delta$ the energy
splitting between the $^2P_{3/2}$ and $^2P_{1/2}$ levels of the iodide
atom. 

\begin{figure}
\centering
\includegraphics[width=0.45\textwidth]{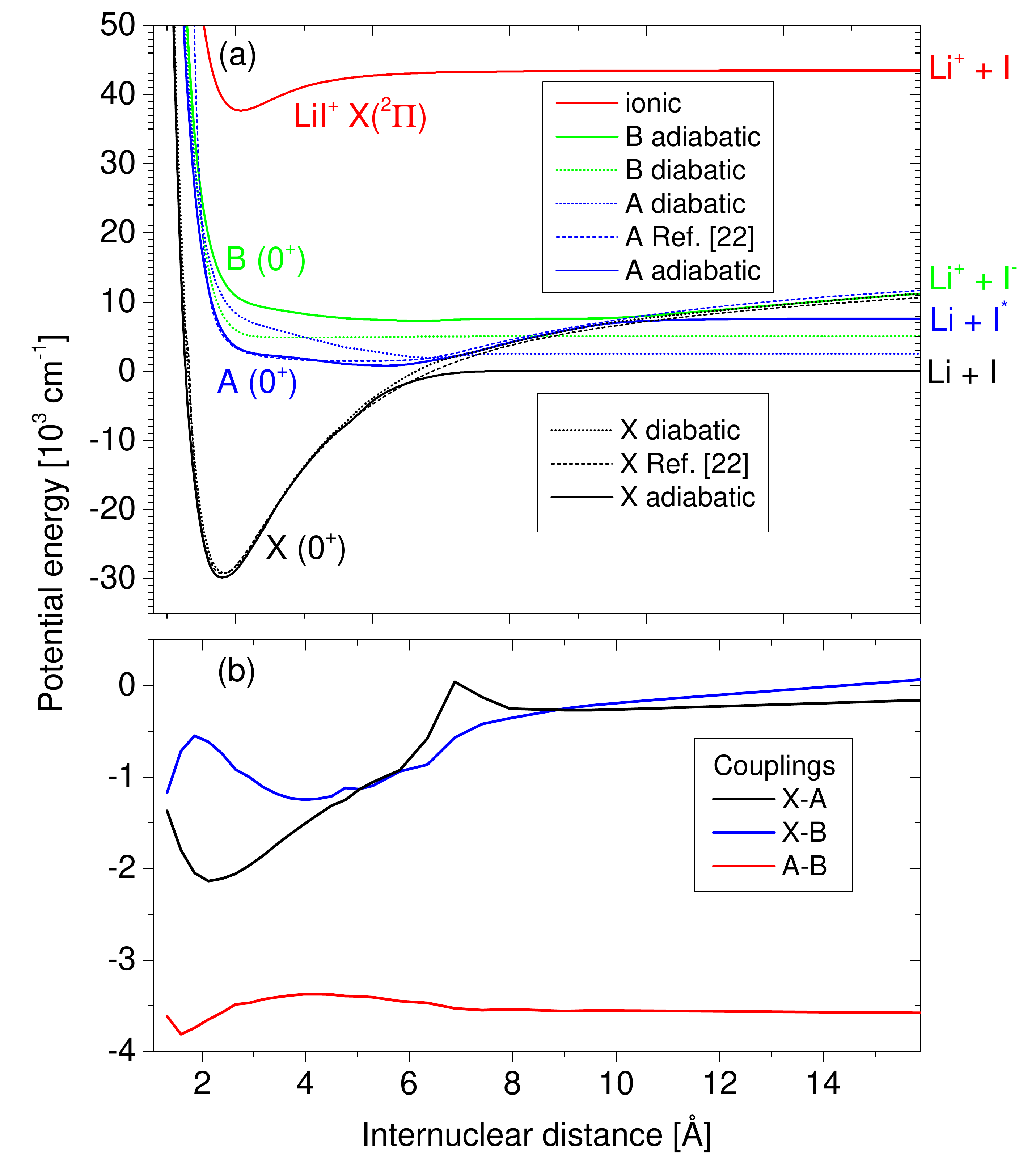} 
\caption{(a) Dashed lines: Diabatic {\it ab initio} potential curves
  calculated in this work (see text for details); Solid lines:
  Adiabatic potentials; Dotted lines: Potentials derived from spectroscopy~\cite{Schaefer:1986}. (b) Diabatic coupling functions.}
\label{fig:Abinitio}
\end{figure}
Figure~\ref{fig:Abinitio}(a) illustrates the  {\it ab initio}
potentials (dashed lines) corresponding to the diagonal entries in
Eq.~\eqref{hampot}. The coupling functions of the three coupled surfaces
given by the off-diagonal elements in Eq.~\eqref{hampot} are shown in
Fig.~\ref{fig:Abinitio}(b). In case of the second excited state, the
potential corresponds to $\frac{1}{3}\Delta+V^{^3\Pi}(R)$. 
Figure~\ref{fig:Abinitio}(a) additionally displays the adiabatic
potentials obtained by diagonalizing Hamiltonian~\eqref{hampot} as
well as the potentials derived from high-resolution spectroscopy~\cite{Schaefer:1986}. One notices a global very
good agreement of the present theoretical results with the
experimental curves. In particular, for the groundstate $X0^+$ the
present theoretical dissociation energy is 29543 cm$^{-1}$ which
compares well with 28839 cm$^{-1}$ from Ref.~\cite{Schaefer:1986}. The
position of the minimum is 2.39~\AA~from theory and 2.40~\AA~from the
experimental fit. For the $A0^+$ state we can compare the $T_e$
parameter by determining the depth of the potential with respect to
the ground vibrational level of the $X$-state. Our theoretical value
of $T_e=30273$ cm$^{-1}$ for the $A0^+$ state compares well with the
experimental result $T_e=30769$ cm$^{-1}$.  

\section{Experiment}
\label{sec:exp}

\subsection{Experimental setup}
\label{subsec:setup}
The experiments are performed using a He nanodroplet beam apparatus
which has been described in detail
previously~\cite{Stienkemeier:2006,Mudrich:2009,Mudrich:2014}. The
main modifications are a time-of-flight detector for measuring ion
mass spectra as well as an amplified femtosecond laser system. The
latter comprises a Ti:Sa-based oscillator (Tsunami by Spectra Physics)
and a regenerative amplifier (Legend by Coherent) operated at 5 kHz
repetition rate. Third harmonic radiation is generated by a home-built
frequency conversion stage which consists of two BBO crystals, a
calcite plate and a $\lambda$/2-wave plate. The laser pulse
characteristics are a pulse energy of up to 20 $\mu$J, a pulse length
of about 230 fs, and a wavelength tunability in the range $\lambda =
260$-$278$ nm. The laser pulses are split into identical pump- and probe pulses which are time-delayed using a mechanical delay line.

An effusive beam of LiI is produced by heating a LiI
sample to 380$^\circ$C using a vapor cell which usually serves as a
doping unit for the He droplet beam. In the experiments with LiI
embedded in He droplets presented in the last section, the LiI cell is
heated to 360$^\circ$C for doping the droplets generated by a
continuous expansion of He out of a 5 $\mu$m nozzle by on average 0.8
LiI molecules. Photoelectron signals are recorded using a standard velocity-map 
imaging spectrometer setup~\cite{Fechner:2012,Vangerow:2014} in time-of-flight detection mode.

\subsection{Gas phase results}
\label{subsec:GasResults}

\begin{figure}
\centering
\includegraphics[width=0.45\textwidth]{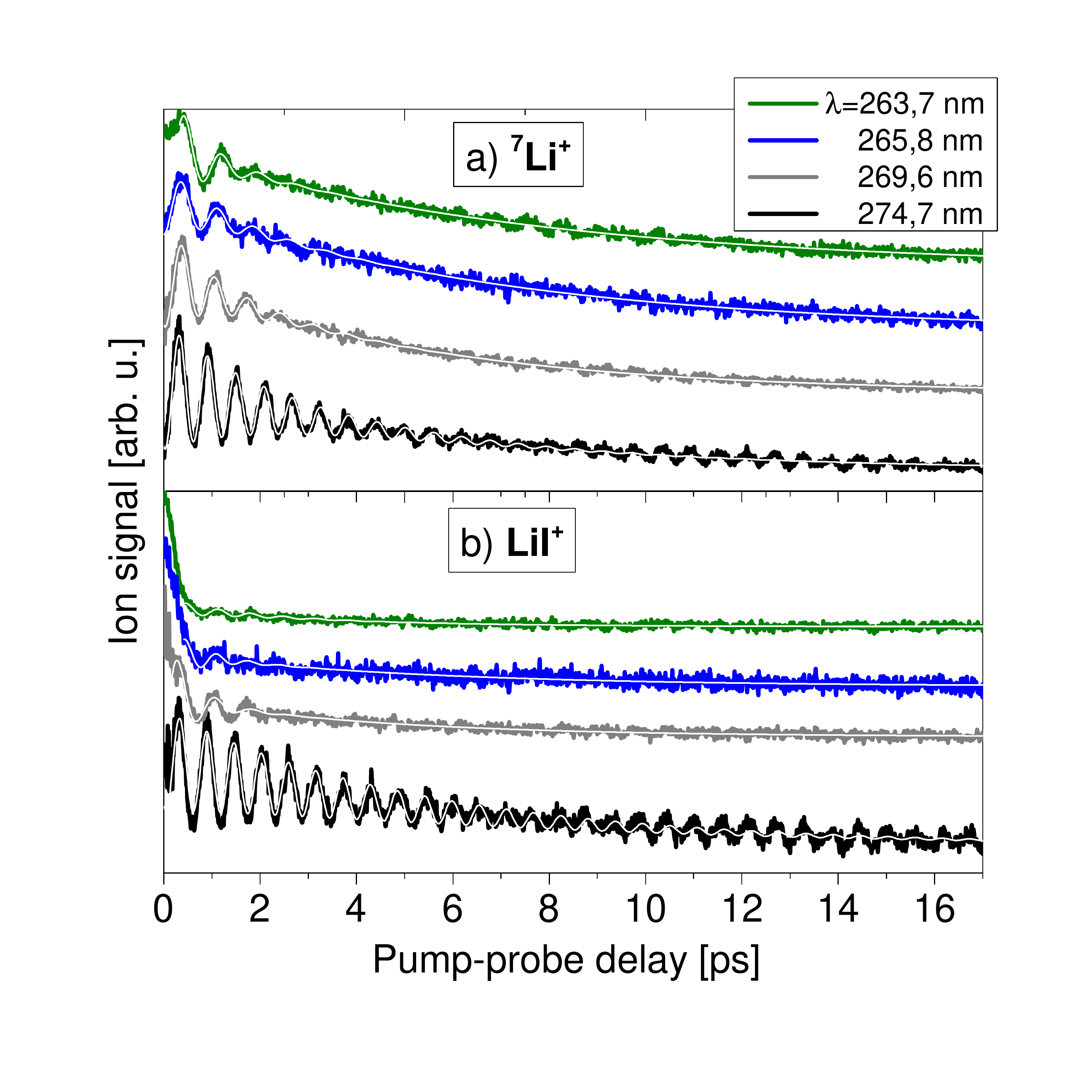}
\caption{Measured pump-probe $^7$Li$^+$ and LiI$^+$-ion signal transients recorded at various center wavelengths of the femtosecond laser. The smooth white lines depict fits to the data.}
\label{fig:traces}
\end{figure}
Figure~\ref{fig:traces} displays the measured Li$^+$ (a) and LiI$^+$ (b)
transient ion signals as a function of the delay time between
pump and probe laser pulses. The traces are vertically shifted from
one another for the sake of visibility. The thin white lines depict
the result of fitting the experimental data to the function  
\begin{equation}
S(t)= S_0+S_1e^{-t/t_1} + S_2e^{-t/t_2}\sin\left( 2\pi (t-t_0)/T\right)\,.
\label{eq:fit}
\end{equation}
It essentially consists of the sum of an exponentially damped
oscillation ($t_2$) to model the wave packet dynamics in the bound state, and
an exponentially decreasing offset ($t_1$) to account for predissociation into
unbound Li and I atoms which elude single-photon ionization. 
We interpret the presence of both LiI$^+$ and Li$^+$ product channels 
by the population of both bound and unbound states close to the dissociation limit of LiI$^+$, respectively, by photodetachment of I$^-$ in the Li$^+$+I$^-$ ion
pair. Experimentally, we find the ratio of Li$^+$ to LiI$^+$ ion
yields to reach a maximum value of about 20 at the laser wavelength
$\lambda =270$ nm.  

The highest contrast of the wave packet oscillation in both LiI$^+$
and Li$^+$ transients is observed at the largest laser wavelength
$\lambda\approx 275$ nm. At that wavelength, damped oscillations up to
about 8 ps are clearly visible. A wave packet is excited deep in the
potential well where dispersion due to the anharmonicity of the
potential curve is minimal. The subsequent reappearance of the
oscillatory signal with maximum amplitude at about 13 ps (``full
revival'') results from partial rephasing of the wave packet and is
determined by the anharmonicity of the
potential~\cite{Averbukh:1989,Vetchinkin:1993,Mudrich:2009}. Moreover, 
at that wavelength the excitation energy falls below the threshold for
predissociation along the $B$-state. The question
whether predissociation via the $B$-state contributes to the reduction
of signal lifetimes will be addressed below.  

\begin{figure}
\centering
\includegraphics[width=0.45\textwidth]{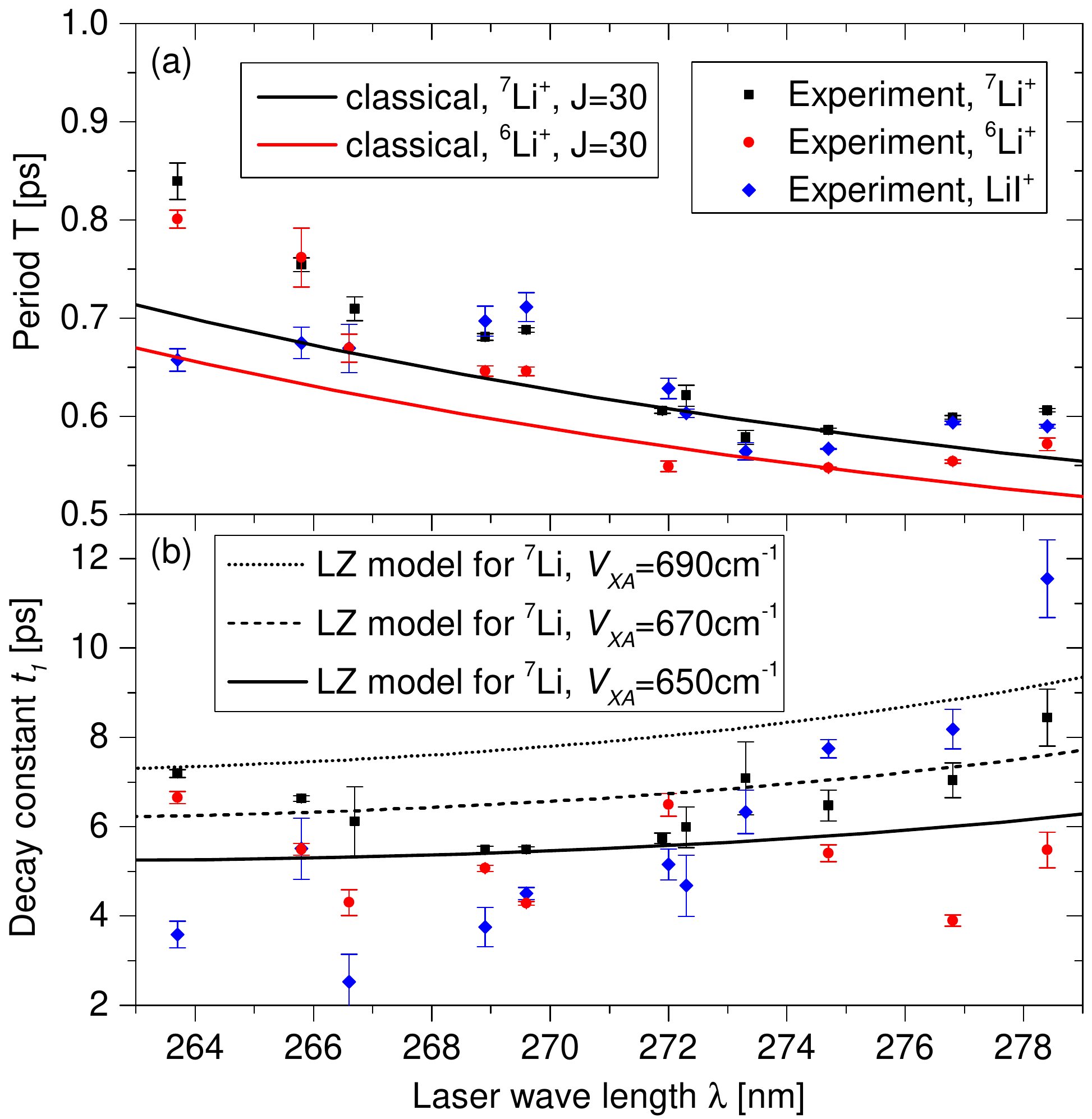} 
\caption{Period (a) and exponential decay time constant (b) of the
  oscillating ion signals obtained from fits of the experimental
  data. The lines in (a) depict the classical oscillation period. The
  lines in (b) show the result of evaluating Landau-Zener's formula
  for the diabatic transition probability at the avoided curve
  crossing between the $X$ and the $A$-states.} 
\label{fig:Ttauclassical}
\end{figure}

For shorter wavelengths $\lambda < 275$ nm, the oscillations quickly
decay. The decreasing contrast is particularly pronounced for the
LiI$^+$ signals, for which the periodic modulation turns into a fast
irreversible drop of the signal within about 0.5 ps when exciting at
$\lambda\lesssim 270$ nm. This decay is assigned to 
direct dissociation via the repulsive diabatic potential of the
$B$ precursor state.
The slow decline of the signal offset on the time scale of several ps
barely changes as a function of $\lambda$. In the following we
concentrate on the discussion of the resulting fit values of the
oscillation period, $T$, and the exponential decay time constant for the
signal offset, $t_1$. These quantities are depicted in Fig.~\ref{fig:Ttauclassical}
a) and b), respectively, as a function of $\lambda$.

As expected for a typical anharmonic molecular potential, the
oscillation periods are found to slightly decrease from about 0.85 to
0.6 ps for increasing $\lambda =264$-$278$ nm, see
Fig.~\ref{fig:Ttauclassical}(a). Despite 
of the considerable scatter of the data points, the oscillation periods
measured for the heavier $^7$Li$^+$ isotope systematically
exceed those of $^6$Li$^+$. No significant deviation of the
oscillation periods measured for Li$^+$ and for LiI$^+$ are found
within the experimental error. 

\subsection{L\MakeLowercase{i}I embedded in He nanodroplets} 
\label{subsec:droplets}
In addition to measurements using effusive LiI, we
report experimental results on LiI embedded in He droplets. The
experimental setup is the same, except that the temperature of the
vapor cell is slightly reduced and the beam of He nanodroplets is
switched on. However, the yield of Li$^+$ ions correlating with the He
droplet beam as compared to the ion signals measured with effusive LiI
is reduced 
by about a factor 100. We attribute this dramatic reduction of the ion yield to the
tendency of atomic ions to form tightly bound complexes (so called
``snowballs'') inside the He droplets due to strong polarization
forces~\cite{Tiggesbaumker:2007}.  

\begin{figure}
\centering
\includegraphics[width=0.42\textwidth]{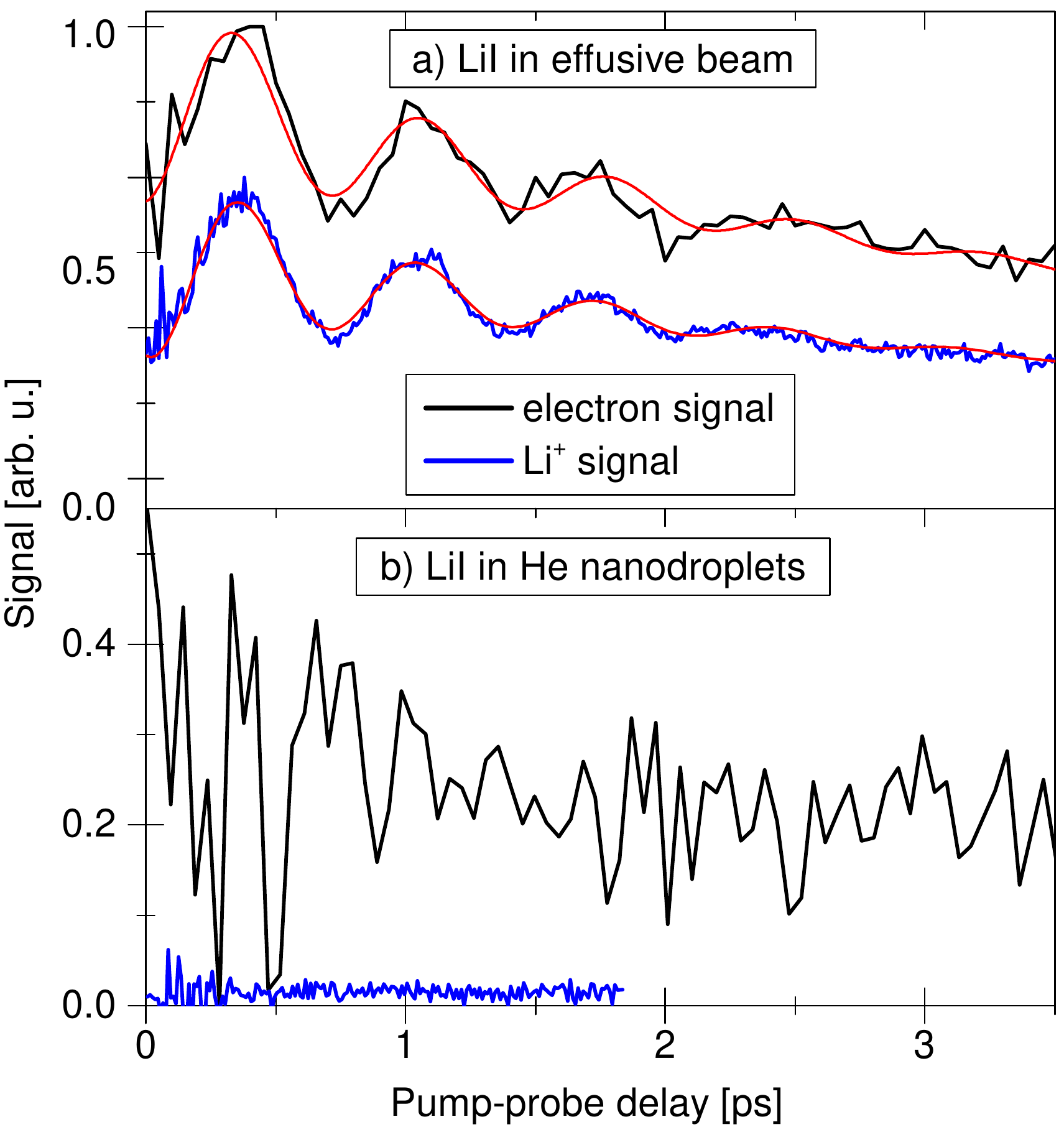}
\caption{Pump-probe transients recorded by measuring the yield of
  $^7$Li ions and of photoelectrons for LiI in an effusive beam (a)
  and for LiI embedded in He nanodroplets (b) at $\lambda=$ 269,6 nm. The smooth lines in (a) are fits to the data.} 
\label{fig:droplet}
\end{figure}
Therefore, we switched to photoelectron detection which we find to be
well-suited for recording the wave packet oscillations for free LiI
molecules, as depicted in Fig.~\ref{fig:droplet} (a). Indeed, the
photoelectron signals for LiI embedded in He nanodroplets are found to
reach about 30\% of the signal level measured with effusive LiI
(Fig.~\ref{fig:droplet} (b)). However, no indications of a periodic
modulation due to wave packet dynamics is observed. Note that we
cannot exclude contributions to the electron signal from
photoionization of contaminations in the He droplets, such as water
picked up by the droplets from the residual gas. The lack of signal
oscillations implies that the coherent vibrational wave packet motion
is efficiently quenched by the interaction with the He
environment. This contrasts previous experiments performed in our group
where the droplets were doped with alkali metal dimers and trimers
which reside in weakly bound dimple-like states at the surface of He
droplets~\cite{Claas:2006,Mudrich:2009,Giese:2011}. In those
experiments, long-lived vibrational coherences were observed by
detecting the neat dopant ions as a function of pump-probe
delay. Thus, molecules immersed in the droplet interior appear to be
subjected to much stronger perturbation of the vibrational dynamics
than those sitting at the droplet surface.

\section{Modelling the predissociation dynamics}
\label{sec:Simulation}

\subsection{Semiclassical calculation}
\label{subsec:semiclass}

To confirm our assignment of the observed dynamics to wave packet
propagation in the $A$-state of LiI we have simulated the vibrational
motion classically for the $A$-state potential $V_A(R)$ augmented by
the centrifugal term 
\begin{equation}
V_C(R)=\hbar^2 J(J+1)/(2\mu R^2). 
\label{eq:centri}
\end{equation}
Here, the
rotational quantum number $J=30$ is used to account for the thermal
population of rotational states at the temperature of the LiI sample
($T_\mathrm{LiI}=653\,$K), and $\mu =6.65$ (5.74) amu 
is the reduced
mass of $^7$LiI ($^6$LiI). The resulting values (solid lines in
Fig.~\ref{fig:Ttauclassical} a)) reproduce the experimental ones
reasonably well. This supports our interpretation of the wavelength
dependence in Sec.~\ref{subsec:GasResults}
in terms of the anharmonicity of the $A$-state potential. 

The decay constant of the signal offset, $t_1$, is modeled using the
standard semi-classical Landau-Zener formula,  
\begin{equation}
P(v_{x})=\exp\left( -\frac{2\pi V_{XA}^2}{\hbar\Delta V_{A} v_{x}}\right)\,,
\end{equation}
for the transition probability from one adiabatic potential ($A$) to another ($X$)
when a wave packet passes an avoided curve crossing. Here $V_{XA}$
denotes the diabatic coupling energy between $X$ and $A$-states,
$\Delta V_A$ is the slope of the difference potential at the curve
crossing, and $v_{x}$ is the classical velocity at the crossing point
in the center-of-mass system. Since in the present scheme the wave
packet passes the $X$-$A$ crossing point twice in each oscillation period, $T$, the
total transition probability for non-adiabatic transition from the $A$
to the $X$-state per period amounts to $2P (1-P)$. This yields 
the time constant for the diabatic transition to the $X$-state, 
\begin{equation}
t_1^{LZ}=\frac{T}{2P (1-P)}.
\end{equation}
It is represented by lines in Fig.~\ref{fig:Ttauclassical}(b)
for $J=30$ and various values of the coupling constant $V_{XA}$. The
Landau-Zener model fits the experimental values well for
$V_{XA}=650(20)$ cm$^{-1}$, reproducing even the trend of a slightly 
increasing $t_1$ for larger values of $\lambda$.   

To assess the contribution to the decay rate of the
second $A$-$B$ avoided curve crossing, we extend the Landau-Zener
model to include the passage at both curve crossings. For the
$A$-$B$ avoided crossing we use $R=10.6\,$\AA~ and
$V_{AB}=5\,$cm$^{-1}$ as obtained for our \textit{ab initio} calculations. The resulting deviation of $t_1^{LZ}$ with
respect to the value obtained by accounting for only the $X$-$A$ avoided crossing
amounts to about 5\%. Thus, the observed signal decay can be
ascribed to predissociation predominantly via the lowest $X$-$A$
avoided crossing. This is in agreement with expectations based on the nearly
perfect diabatic behaviour of the reconstructed potential curves at the outer $A$-$B$
crossing~\cite{Schaefer:1986}.

\subsection{Quantum dynamics calculations}
\label{subsec:qudyn}

A more realistic modeling of the detected wave packet dynamics is
achieved by treating the dynamics fully quantum mechanically including
couplings between the relevant diabatic states and with the laser
fields. In particular our goal is to ascertain the role in the
predissociation dynamics of the second $A$-$B$ curve crossing. 
To model the probe  process in the presence of the laser field $E(t)$,
the Hamiltonian~\eqref{hampot} is extended   by adding the 
potential for the groundstate of the LiI$^+$ ion and the dipole
couplings $ d \cdot E(t)$ between the two $^1\Sigma$ states of the
LiI molecule and the ionic state. The dipole transition moments $d$
are assumed to be constant.

The simulations employ the first three coupled diabatic LiI potentials
and their coupling functions, as expressed by
Hamiltonian~(\ref{hampot}), as well as the LiI$^+$ ionic groundstate
potential. We assume that the observed wave packet dynamics involves  
exclusively $0^+$ potentials. Initial pumping into the
$\Omega=1$ states can be neglected due to the small transition dipole
moment and the repulsive character of these electronic states. 
This approximation is justified also by the fact that in spectroscopic 
observations of the transition $X0^+ \to A0^+$ there was no indication         
of perturbations coming from the $\Omega=1$ states~\cite{Schaefer:1986}.
To account for the thermally excited molecules, a rotational barrier
parametrized by $J$ is added to each potential according to Eq. (\ref{eq:centri}). 

The pump process is simulated by defining a Gaussian shaped vibrational wave packet in the first excited state at $t=0$. The peak position and width are determined by mapping the groundstate vibrational wave function in the $X$-state onto the $A$-state potential according to the spectral intensity distribution of the laser pulses used in the experiment and the $V_A(R) - V_X(R)$ difference potential. 
For simplicity, only the vibrational groundstate is assumed to be populated prior to the excitation step. Thus, the calculations include propagation of the initial wave packet and 
interaction with the time delayed probe
pulse. Consequently, the population transferred to the ionic state via
single-photon absorption is calculated as a function of delay time. 
Since the rotating wave approximation cannot be applied in these
calculations, 
the time grid is required to have a resolution of $97\,$as to
accurately sample the oscillating electric field of the ultraviolet
laser pulses. The 
emission of photoelectrons upon excitation into the ionic state is not
taken into account. 

Due to the predissociative character of the first two excited states,
artificial reflections of the wave packet at the edge of the numerical
grid occur which are suppressed by adding complex absorbing
potentials. Following Ref.~\cite{Riss:1993}, we use a quadratic
complex potential,   
\begin{equation}\label{eq:CAP}
V(r)= -i\eta\cdot\theta(R-R_c)\cdot(R-R_c)^2\,,
\end{equation}
for the dissociative first two excited states, for which 
reflection plays a role. Equation~\eqref{eq:CAP} employs the 
Heaviside step function $\theta$ and is characterized by the
parameters $\eta$ (strength of the potential) and $R_c$ (onset of the
potential). $\eta$ and $R_c$ need to be chosen such that 
all dissociating fractions of the wave function at a
sufficiently large distance are absorbed.

\begin{figure}
\centering
\includegraphics[width=0.45\textwidth]{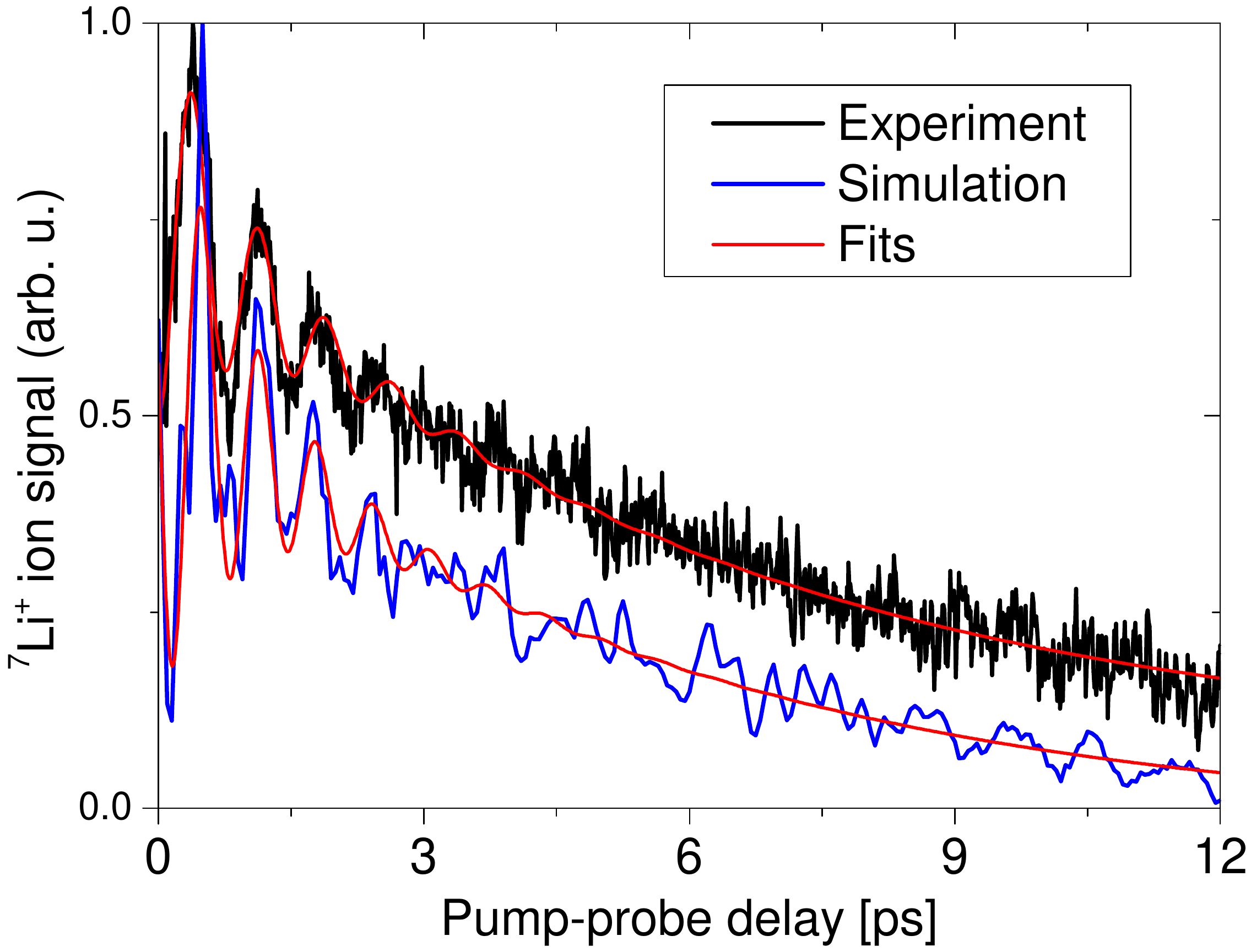}
\caption{Experimental and simulated pump-probe transient $^7$Li$^+$ ion
  yield at the center wavelength $\lambda = 266.7$ nm. The smooth
  lines are fits of expression (\ref{eq:fit}).} 
\label{fig:Simtraces}
\end{figure}
The time-dependent Schr\"odinger equation is solved by expanding the
time evolution operator in Newton polynomials~\cite{Kosloff:1994}.
The pulse parameters are chosen to match
the experiment in that the pulse length (full width at half maximum,
FWHM) $t_p=230$ fs is determined by an autocorrelation measurement
whereas the intensity of the pulses in the focal region is estimated to
$I\approx 2\times 10^{11}$ W/cm$^2$ based on the measured beam
diameter. For a
detailed comparison with the experiment, the resulting transient
signals for rotational states up to $J=70$ have to be thermally
averaged according to the Boltzmann distribution of LiI at $T=673$ K, 
\begin{align}
\langle S \rangle=\sum_J  S_J g_J P_J/\sum_J g_J P_J.
\end{align}
Here, $P_J=\exp\left( -E_J/k_\mathrm{B}T\right)$, $E_J=hcB_0J(J+1)$
denotes the Boltzmann factor, $g_J=2J+1$ is the degeneracy of the
$J$-th rotational level, and $S_J$ is the simulated delay time
dependent population in the ionic state for given $J$. A comparison of
the pump-probe traces at the wavelength $\lambda = 266.7$ nm of the
experiment and a simulation for which every second rotational state
between 6 $\le J \le$ 70 is taken into account for simplification is shown in
Fig.~\ref{fig:Simtraces}. The traces show a good agreement in terms of
the period and damping of the oscillations as well as the exponential
decay. Also, the vibrational revival occurs at a similar time
of roughly 10$\,$ps.  

\begin{figure}
 \centering
  \includegraphics[width=0.45\textwidth]{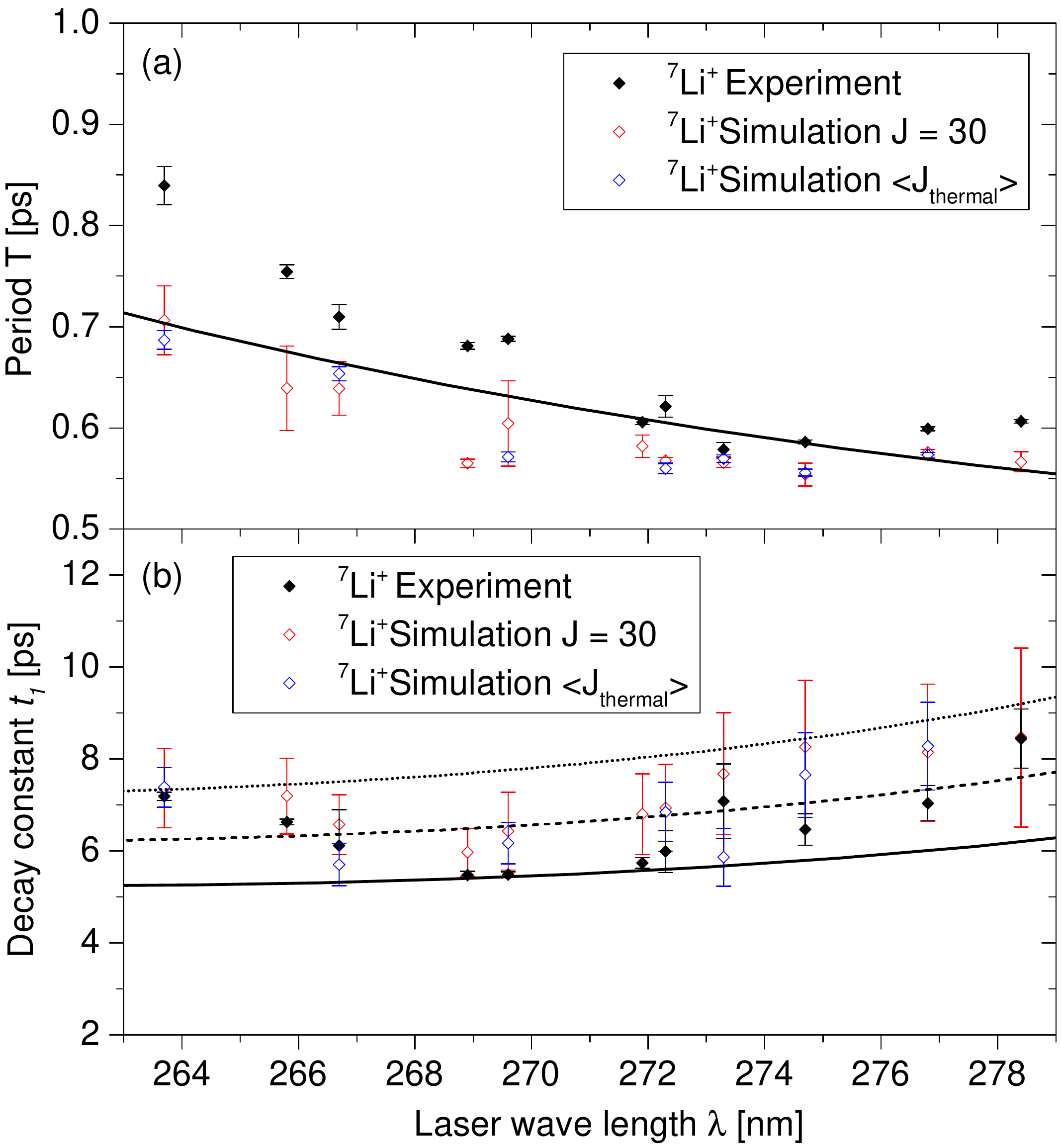} 
  \caption{Fitted experimental (filled symbols) and simulated (open
    symbols) oscillation periods (a) and exponential decay time
    constants (b). Solid lines represent the classical oscillation periods (a) and the Landau-Zener model (b) as in Fig.~\ref{fig:Ttauclassical}.} 
\label{fig:Ttauquantum}
\end{figure}

For further simplification, the calculations for the
wavelengths investigated in the experiment are performed with a
single state, $J = 30$, only. The simulated pump-probe transients are analyzed in the same way as the experimental data by fitting model Eq.~(\ref{eq:fit}). The resulting simulated oscillation periods $T$ and decay constants $\tau_1$ are shown in
Fig.~\ref{fig:Ttauquantum} where they are compared to the
experimental values. The agreement between the
simulations and the experiment is fairly good, confirming the
accuracy of the {\it ab initio} potentials and couplings. The
simulated values of $T$ shows the same dependence on the
wavelength. The decay constant $\tau_1$ becomes minimal for $\lambda\approx
269\,$nm as in the experiment. However, slight differences between
simulation and experiment are present. In particular $T$ lies 
systematically below the measured values. This can be
attributed to neglecting any initial vibrational excitation of LiI in
the quantum dynamics simulations.
The results of simulations, where full thermal averaging over
rotational states is performed, are depicted in
Fig.~\ref{fig:Ttauquantum} as well. Thermal averaging does
not significantly improve the agreement with the experiment which justifies
the simplified approach using a single state, $J = 30$.  

\section{Summary}
\label{sec:summ}

We have studied the predissociation dynamics of LiI in a one-color
femtosecond pump-probe experiment  for molecules in the gas phase
and embedded in He nanodroplets. The wavelength was varied in the range
$\lambda=260\,$-$278\,$nm. We have corroborated the experimental
observations by full quantum dynamics simulations based on newly
calculated \textit{ab initio} potential energy curves and
non-adiabatic couplings. Vibrational wave packet 
oscillations in the predissociative $A$-state are observed in both ion
and electron yields for gas phase LiI molecules. The exponential
decay of the signal with increasing pump-probe delay times is
interpreted in terms of predissociation due to diabatic coupling
between the excited $A$-state and the ground  $X$-state. 
The good agreement of the experimental and simulated pump-probe
transients confirm the accuracy of the calculated potentials and
coupling energies.  Predissociation is found to occur nearly
exclusively  via the $X$-$A$ avoided curve crossing. Since it is a
single crossing that determines the predissociation, the decay
constant as a function of laser wavelength can also be
reproduced by a semiclassical Landau-Zener model for the diabatic
transition. The reason that the dynamics is dominated by the
$X$-$A$ avoided crossing is due to the strong diabatic coupling
between these states, found to amount to 650(20)$\,$cm$^{-1}$. This is
much larger than the coupling between the 
$A$-state and the $B$-state of only 5(2)$\,$cm$^{-1}$.

Analogous measurements with LiI embedded in He
nanodroplets result in photoion signals that are strongly suppressed
compared to the gas phase measurements. This is attributed to
complex formation inside the droplets upon ionization. While the photoelectron signals have comparable
amplitudes to those measured for free LiI, no wave packet oscillations
are observed. This may be explained by a strong coupling of the
excited LiI to the He droplet which induces fast damping of the
coherent vibrational wave packet motion. 

Our observation is in contrast to femtosecond pump-probe experiments
with diatomic alkali metal molecules attached to He nanodroplets
for which the vibrational motion is well
resolved~\cite{Claas:2006,Claas:2007,Schlesinger:2010,Mudrich:2009,Gruner:2011}. The 
difference is rationalized by the different coupling of the molecules
to the droplet environment. It results from different locations,
on the droplet surface for the weakly coupled alkali dimers whereas 
molecules immersed in the droplet interior such as LiI are subjected to
strong coupling. For molecules inside the droplets, a similarly
strong effect of the He environment on the molecular 
dynamics as observed here has also been reported for recent impulsive alignment
experiments~\cite{PentlehnerPRL:2013}. The
droplets were found to alter the rotational period and suppress
alignment revivals. 

These findings together call for a more comprehensive study of the ultrafast
dynamics of molecules immersed in a He quantum fluid. For our
pump-probe experiment, refined
detection schemes such as ion imaging techniques would allow to
unequivocally confirm suppression of coherent wave packet
motion. More generally, it would also be interesting to see whether the regime of
intermediate coupling between molecule and He droplet can be
explored by using different molecular species, \textit{e.\,g.} alkali-earth alkaline mixtures~\cite{Lackner:2014}. This would provide much needed input for the theoretical modeling of the interaction of molecules with He droplets and thus be crucial for
developing a rigorous understanding of quantum fluid environments.

\begin{acknowledgments}
  The authors wish to thank E. Tiemann for providing the LiI $X$ and
  $A$-state potential energy curves derived from spectroscopy, and
  W. Strunz for fruitful discussions. We would like to acknowledge the use of the computing resources provided by bwGRiD (http://www.bw-grid.de), member of the German D-Grid initiative. Financial support 
  by the Deutsche Forschungsgemeinschaft, the Alexander von Humboldt foundation (WS), and the Russian
  Foundation of Basic Research (ASB, AVB) is gratefully acknowledged.
\end{acknowledgments}



\end{document}